\documentclass[aps,prd,reprint,floatfix,nofootinbib]{revtex4-2}

\providecommand{\planss}{Planet. Space Sci.}
\providecommand{\aj}{Astron. J.}

\usepackage{amsmath}
\usepackage{amssymb}

\usepackage{hyperref}
\usepackage{xcolor}
\usepackage{graphicx}

\makeatletter
\protected\def\unit{\@ifstar\@unitns\@unit}
\newcommand{\@unit}[1]{\ensuremath{\ \mathrm{#1}}}
\newcommand{\@unitns}[1]{\ensuremath{\mathrm{#1}}}
\makeatother

\newcommand{\pd}[2]{\ensuremath\frac{\partial #1}{\partial #2}}
\newcommand{\expect}[1]{\ensuremath\left\langle #1\right\rangle}

\newcommand{\Sun}{\ensuremath\odot}
\newcommand{\Earth}{\ensuremath\oplus}

\begin{document}
	
	\title{Direct detection signatures of a primordial Solar dark matter halo}
	
	\author{Noah B.~Anderson}
	\affiliation{Department of Physics, Truman State University, Kirksville, MO 63501, USA}
	
	\author{Angelina Partenheimer}
	\affiliation{Department of Physics, Truman State University, Kirksville, MO 63501, USA}
	
	\author{Timothy D.~Wiser}
	\email{tdwiser@truman.edu}
	\affiliation{Department of Physics, Truman State University, Kirksville, MO 63501, USA}

	\date{\today}

	\begin{abstract}
		A small admixture of dark matter gravitationally bound to the proto-Solar gas cloud could be adiabatically contracted into Earth-crossing orbits with a local density comparable to (or even exceeding) the Galactic halo density. We show that a significant fraction ($\sim 25\%$) of the resulting `Solar halo' would remain today, surviving perturbations from Jupiter and close encounters with Earth, and would be potentially observable in direct detection experiments. The population would have distinct signatures, including a nonstandard annual modulation and extremely low velocity dispersion compared with the Galactic halo, making it an especially interesting target for coherent or resonant detection of ultralight particles such as axions or dark photons.
	\end{abstract}
		
	\maketitle
	
	
	\section{Introduction\label{sec:introduction}}
	
	Dark matter (DM) makes up a majority of the non-relativistic matter in the Universe~\cite{history_dm,planck_2018_params}, but its interactions---other than gravitational---are unknown. Candidate particle physics models for DM include weakly interacting massive particles (WIMPs)~\cite{wimp_review}; primordial black holes~\cite{pbh_review} or other massive compact halo objects (MACHOs); and various ultralight bosons~\cite{uldm_review}, such as axions~\cite{axion_cosmology}. As experimental constraints on WIMPs and MACHOs become more severe~\cite{lux_results,xenon_results,macho_lensing}, ultralight DM candidates have come under renewed interest. Several direct detection experiments focused on ultralight DM are in development or underway~\cite{graham_review,admx_latest,casper_proposal,dm_radio}.
	
\begin{figure}
			\includegraphics{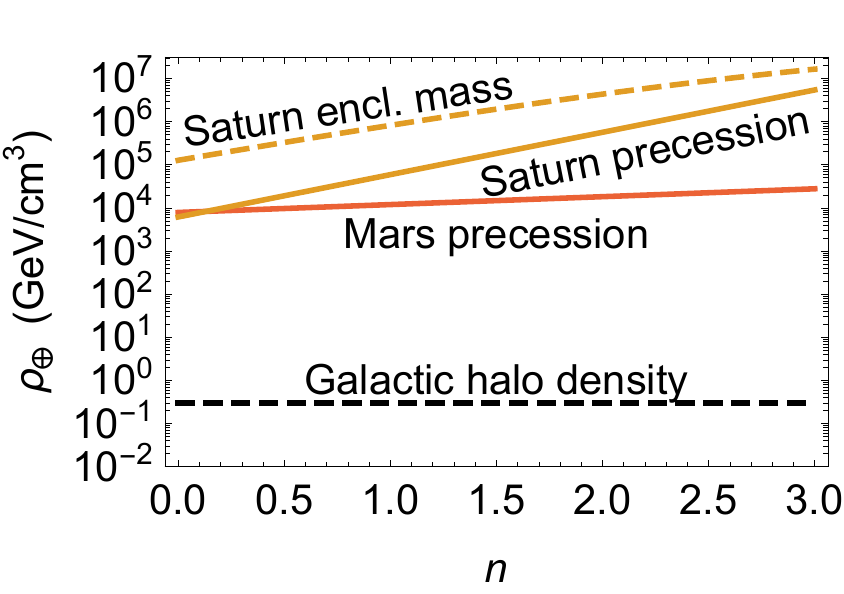}
		
			\caption{\label{fig:densitylimits}Upper bounds on $\rho_\Earth$, the density at Earth of a Solar-bound DM population. These limits are derived from planetary motion~\protect\cite{Pitjev:2013sfa} and depend only on the gravitational effect of the DM, assuming a spherical density profile with $\rho(r)\propto r^{-n}$. The precession of Saturn's perihelion (gold, solid) sets the strongest limit on a uniform density ($n=0$) but for cuspier profiles, Mars (red, solid) sets a stronger limit on the density at Earth due to their relative proximity. The precession limits are sensitive to the local density at a particular planet and are extrapolated back to Earth's position. Saturn's orbital period (gold, dashed) also sets an independent limit (from Kepler's 3rd Law) on the total enclosed DM mass between Earth and Saturn. Although a weaker limit, it is less sensitive to local variations in the DM density (for instance, if a planet cleared it orbital region of DM). The local density of DM from the Milky Way halo (black, dashed) is shown for comparison only; it is not a limit on the Solar halo density.}
		\end{figure}

	Currently, the smallest scale on which DM has been detected is the galactic scale, by the measurement of rotation curves in other galaxies~\cite{rubin_andromeda,rotation_curves_things} and our own Milky Way~\cite{milky_way_dm}. Below the scale of galaxies, baryonic matter dominates both gravitational and nongravitational interactions. It is an open question whether DM does have (or even can have) detectable effects on smaller scales, particularly within our own Solar System. The possibility of DM in the Solar System has been explored along several axes, though most work assumed that (1) DM is composed of WIMPs and (2) DM in the Solar System is captured directly from the local region of the virialized Galactic halo. While these assumptions make studies more predictive, they are significant limitations. Interestingly, the assumption of weak-scale DM--nuclear scattering cross sections has a substantial effect on the lifetime of DM trapped in the Solar System, as it undergoes significant scattering within the Sun and planets over billions of years. Ultralight DM candidates typically have much smaller cross sections and, therefore, may remain bound for longer. And, while Solar halo DM captured directly from the Galactic halo typically has a lower density than the Galactic halo itself, even a tiny amount of DM that was gravitationally bound to the proto-Sun can be `amplified' to detectable levels---even exceeding that of the Galactic halo---via adiabatic contraction. The amount of Solar halo DM is constrained (purely gravitationally) by observations of planetary motion and comparisons to computed ephemerides~\cite{Pitjev:2013sfa}. The resulting limits on the Solar halo density at Earth are shown in Fig.~\ref{fig:densitylimits}; the upper bounds are orders of magnitude above the Galactic halo density.
		
	As the Sun formed through gravitational collapse, the gravitational potential created by the baryons changed slowly, becoming more and more tightly binding. When such collapse is slow, the fast-moving DM particles of the Galactic halo are not efficiently captured, since the gravitational potential does not change much during one crossing time. On the other hand, particles that are already gravitationally bound experience the full effect of the collapse, and can be drawn significantly closer to the barycenter of the region: the newly forming Sun. This process---adiabatic contraction---generically results in an enhancement of the density close to the Sun and a corresponding increase in the velocity dispersion (by Liouville's Theorem). Because the Earth is relatively close to the Sun (when compared to the initial gas cloud radius of $\sim\unit*{ly}$), a small initial density of bound DM can be greatly enhanced near Earth. We call such a population ``Solar halo DM,'' in constrast to the better-established ``Galactic halo DM.''
	
	Adiabatic contraction of DM around forming stars is not a new idea: It has been studied in the context of Population III stars capturing enough DM to alter their structures (``dark stars''~\cite{dark_stars_review}) and as a way to set limits on star-destroying DM such as primordial black holes~\cite{adiabatic_pbh}. In the Solar System, we know (from planetary motions~\cite{Pitjev:2013sfa}) that the DM self-gravity is negligible, and we assume (because we are motivated by ultra-weakly interacting types of DM) that the nongravitational interactions are unimportant, so unlike previous work we focus on the detectability of this population of DM \emph{at Earth}. The focus on DM in Earth-crossing orbits, as well as a presumed origin during the Solar System's formation, raises new challenges, particularly the longevity of DM orbits in the presence of perturbations from the planets over billions of years.
	
	The most compelling DM candidates for detection as part of a Solar halo are ultralight bosonic candidates that can be detected via coherent oscillations, including axions, axion-like particles (ALPs), dark photons, and dilatons with masses between a few tens of \unit*{feV} and a few \unit*{\mu eV}. The lower bound on the mass comes from the requirement that the de Broglie wavelength of a Solar DM particle `fit' inside Earth's orbit so that adiabatic contraction can be effective and so that applying particle mechanics to the DM trajectories is sensible, while the upper bound comes from practical design requirements of coherent or resonant detection experiments. The detectability of ultralight DM candidates in this frequency range is, in many cases, limited by the width of the signal in frequency space. In contrast to WIMP--nucleon scattering, ultralight DM with a lower velocity dispersion typically deposits \emph{more} power in a detector by enhancing resonant build-up and/or reducing the relevant bandwidth of background noise. Additionally, ultralight DM candidates typically (though not universally) have much smaller scattering cross sections with normal matter, which prevents them from being strongly absorbed in the Sun or Earth during their time in the Solar System (the ultimate fate of Solar System-bound WIMPs with weak-scale cross sections~\cite{edsjo_diffusion}). However, much of our work applies equally well to non-ultralight candidates, including heavier ALPs and even WIMPs with sufficiently small nuclear cross sections.
	
	Our primary motivation in this work is to establish that DM bound to the Sun is (1) plausibly present and detectable today and (2) has distinct observational signatures which may motivate new searches and/or reanalyses in existing direct detection experiments. So far we have not been able to determine a compelling lower bound on the density of Solar halo DM, and so we do not attempt to set any new limits on the viability of dark matter candidates from non-observation in existing direct detection experiments. (Existing data \emph{can} constrain the density of Solar halo DM for certain DM candidates and mass ranges, which could exclude DM candidates only if future work sets some expectation for the minimum Solar halo density.) Instead we focus primarily on the observational properties of such a halo, with the goal of providing a novel discovery or confirmatory channel for certain DM candidates.
	
	While we cannot yet set a firm lower bound on the density of the Solar DM halo, we do want to establish some sense of scale for what would be required for a detectable halo. Suppose that we want the density of Solar halo DM at Earth to exceed the Galactic halo density---that would make the signal definitely detectable in most bosonic DM direct detection experiments (in coherent or single-particle absorption channels). As we will show in Sec.~\ref{sec:distribution}, a pessimistic assumption about the initial bound DM distribution function leads to a power-law density profile $\rho(r)\propto r^{-3/2}$ after adiabatic contraction. The total mass of the contracted profile is then on the order of $10^{-10}M_\Sun$; that mass would have initially been spread almost-uniformly across the proto-Solar gas cloud, a nearly negligible `contamination.' Conversely, for the Solar DM halo to end up \emph{subdominant} to the Galactic halo after adiabatic contraction, the processes separating DM from baryons throughout the structure formation process must have been effective at reducing the DM concentration by a factor of at least $10^{10}$.
	
	In Sec.~\ref{sec:distribution}, we will characterize the properties of orbits for adiabatically contracted DM, bracketing the final distribution function between two extreme initial conditions. The detailed properties of the final distribution function, and therefore the details of the initial distribution function, are not essential for our conclusions, but give a useful foundation for discussing the 4.6 billion years to come. Sections~\ref{sec:jupiter} and~\ref{sec:earthcrossing} establish the longevity of orbits against perturbations from the planets; for normal matter, these perturbations are highly effective at clearing out orbits in the inner Solar System, but for DM, they are only partially effective. The combined actions of Jupiter, Earth, and Venus clear out $\sim 75\%$ of Earth-crossing orbits by today, leaving a significant fraction for detection. In Sec.~\ref{sec:detection} we discuss the distinct observable properties of a Solar DM halo, and we conclude in Sec.~\ref{sec:discussion}.
	
	In this paper we will move frequently between thinking of orbits in position- and velocity-based phase space and in terms of their Keplerian orbital elements, depending on which is appropriate. Appendix~\ref{app:elements} reviews our notation for the orbital elements as well as some simple but important relations for determining orbital intersections.

	\section{Initial Distribution\label{sec:distribution}}
	

	In this section we discuss the phase space distribution function resulting from adiabatic contraction of an initial population of DM during the formation of the Sun. We do not assume any particular value for the mass fraction of DM bound to the molecular cloud (MC) soon to become the Sun, except that it is probably not zero. While DM is generically `left behind' by cooling and collapsing baryons, some slower-moving particles will remain trapped in the collapsing cloud. Steigman et al.~\cite{steigman} studied several mechanisms for the purely gravitational trapping of DM within collapsing baryons. While it is inefficient for a slowly-collapsing cloud to trap fast-moving Galactic DM, there may be yet other ways for DM to become bound within MCs. The early generations of stars likely formed in dense DM halos and may have contained significant DM mass fractions~\cite{dark_stars_review}; the resulting supernovae seeded the Galaxy with metals, up to a level of $M_{\rm metal}/M_\Sun \sim 0.01$ in our own Sun. Of course, metals have drastically different interactions than DM, so we cannot claim that a proportional amount of DM would be found orbiting the Sun. Instead we simply note that tiny initial mass fractions can become non-negligible densities at Earth after the adiabatic contraction process. In the virial initial condition scenario described below, it takes a DM mass fraction of only $10^{-10}$ to, after adiabatic contraction, match the local Galactic halo density at Earth. As a result, even extremely inefficient methods for getting DM bound within MCs may prove to be detectable.
	
	It is appropriate to ask if the Sun's formation is really an adiabatic process from the viewpoint of the DM. The relevant timescale is the DM orbital period, which is roughly the freefall timescale $t_{ff}$ of the gas cloud. Note that the collapse can never proceed \emph{faster} than the freefall time, so the sudden approximation (in which a DM particle is assumed to have fixed position and velocity during the collapse) is not particularly appropriate. Furthermore, the collapse need not be parametrically \emph{faster} than the freefall time; Ref.~\cite{adiabatic_pbh} checked using numerical simulations that a collapse time $t_c\gtrsim 3t_{ff}$ is sufficient for the adiabatic approximation to be accurate. It may be the case that some phases of the Sun's formation history were rapid enough to reduce the accuracy of the adiabatic approximation. There may also be reductions in accuracy from inhomogeneity or asphericity of the collapsing baryons. In our view, especially considering the vast uncertainty of the initially bound DM abundance, the adiabatic calculation provides an easily calculable and physically plausible scenario. A highly detailed understanding of the gravitational potential of the forming Sun, including timescales, geometry, and clumpiness, over many scales, would be necessary to do better, and we leave it for future work. None of our conclusions depend on the actual density present at Earth, and only one minor result (Eq.~\ref{eq:densityvariation}) will depend explicitly on the shape of the density profile.
	
	With the caveats of the previous paragraph, we assume a simple model of the Sun's formation: a uniform sphere of gas with radius $R_i\sim\unit*{ly}$ and mass $M_\Sun$ which collapses slowly. Since the DM is essentially collisionless and subdominant to the mass of the Sun, we treat the DM as test particles and we can use Liouville's theorem to relate the initial and final phase space density simply by following along the trajectory of the particles during the collapse. Assuming the initial phase space density of DM is time-independent and the gas collapse remains adiabatic, there are adiabatic invariants that completely determine the final phase space density. Since the form of the Hamiltonian changes over time, but begins and ends with spherical symmetry, we use the action variables in spherical coordinates $(J_r,J_\theta,J_\phi)$, defined by $J_i=\oint p_i dq_i \textrm{ (no sum)}$, as our adiabatic invariants. (Explicit formulas for the adiabatic invariants are listed in Appendix~\ref{app:invariants}.) We therefore compute the final phase space density as follows:
	\begin{itemize}
		\item For a point in the final phase space, compute the adiabatic invariants $(J_r,J_\theta,J_\phi)$ using the final (Keplerian) Hamiltonian with $V(r)=-\frac{GM_\Sun}{r}$;
		\item find a corresponding point in the initial phase space that has the same adiabatic invariants under the initial (simple harmonic) Hamiltonian with $V(r)=+\frac{1}{2}\frac{GM_\Sun}{R_i^3}r^2$;
		\item and evaluate the initial phase space density at that point.
	\end{itemize}

	It is illustrative to consider the simple case of a circular orbit. A circular Keplerian orbit has $J_r=0$ and $J_\theta+J_\phi=2\pi L$ where $L$ is the specific angular momentum. In the initial Hamiltonian, these adiabatic invariants require a circular orbit with the same angular momentum. The radii and velocities of the orbits are therefore related:
	\begin{align}
		r_f v_f &= r_i v_i \nonumber\\
		r_f \sqrt{\frac{GM_\Sun}{r_f}} &= r_i \sqrt{\frac{GM_\Sun}{R_i^3}r_i^2} \nonumber\\
		r_f &= \frac{r_i^4}{R_i^3}. \label{eq:radius_relation}
	\end{align}
	In the above equations, $r_{i,f}$ and $v_{i,f}$ refer to the orbital radius and velocity before and after the contraction, respectively; $R_i$ is, as above, the initial radius of the gas cloud that eventually forms the Sun. Per Eq.~\ref{eq:radius_relation}, when $r_i \ll R_i$, $r_f\ll r_i$ and the final orbit is significantly contracted. As a result, the DM density sharply peaks near the Sun.
	
	If all orbits are initially circular, Eq.~\ref{eq:radius_relation} is sufficient to determine the final density in terms of the initial density:
	\begin{equation}
		\rho_f(r) = \frac{1}{4}\rho_i(r_i)\left(\frac{r}{R_i}\right)^{-9/4}. \label{eq:circular_density}
	\end{equation}
	The resulting $r^{-9/4}$ power law (from an initially uniform density $\rho_i$) matches the assumptions and result of the adiabatic contraction calculation in Ref.~\cite{steigman}.
	
	However, if the initial DM orbits are not circular, the density profile tends to be shallower. We consider the extreme opposite case: an initial Boltzmann distribution, which has a homogeneous, isotropic Maxwellian velocity distribution with width $\sigma_v$ and a Gaussian density profile with width $\sigma_r=\sqrt{R_i^3/GM_\Sun}\sigma_v$. (This relation between $\sigma_r$ and $\sigma_v$ ensures a time-independent phase space density.) When the radius of the DM cloud is on the order of the gas cloud ($\sigma_r\sim R_i$), $\sigma_v$ is on the order of escape velocity, so it can not be much larger. We call this initial distribution the `virial' case; even though the DM cloud is not self-gravitating, its radius is supported by its (homogeneous) velocity dispersion. In the virial case we follow the above procedure for computing the adiabatic invariants and obtain a density profile with an approximate $r^{-3/2}$ power law at small radii.
	
	The exponent of this power law can be understood heuristically by estimating the number of particles that always remain within a final radius $r$, $N(r)$, by noting that those particles must be simultaneously inside a fixed initial radius $r_i\propto r^{1/4}$ (by analogy with Eq.~\ref{eq:radius_relation}) \emph{and} below a cutoff speed $v_i\propto r_i$ (to avoid going too far outside of $r_i$ in the initially simple-harmonic potential). Since the inital phase space distribution approaches a constant for small $r_i$ and $v_i$, $N(r)$ is just proportional to the available phase space volume. Therefore $N(r)\propto r_i^3 v_i^3\propto r^{3/2}$, and
	\begin{equation}
		\rho_f(r) \simeq \frac{1}{2}\rho_i(r_i)\left(\frac{r}{R_i}\right)^{-3/2} \label{eq:virial_density}
	\end{equation}
	at $r$ small enough that $r_i \ll R_i$. This power law matches the one obtained numerically, with similar assumptions and initial conditions, by Ref.~\cite{adiabatic_pbh}.
	
	In both the circular and virial initial distributions, the density is sharply enhanced by a power of $R_i/r$ at small radii after adiabatic contraction, and the final orbital speeds are on the typical virial scale $\sim\sqrt{GM_\Sun/r}$. However, the character of the final distributions is significantly different. Consider Earth-crossing orbits. In the circular case, all the final orbits are also circular and therefore all have identical speed in the Sun's frame. In the virial case, there is a wide distribution of speeds and directions; the Earth-crossing orbits have semimajor axes as low as $0.5\unit{AU}$, ranging to arbitrarily high. These properties are relevant for the direct detection of the particles, to be discussed in Sec.~\ref{sec:detection}.
	
	For the rest of this paper we will focus on the virial initial distribution solely because it generally results in more conservative prospects for detectability: a smaller final density at Earth, a wide range of semimajor axes (resulting in a wide variety of planetary perturbations), and a broader distribution of speeds. Whatever the initial distribution of bound DM, the final distribution likely interpolates between the highly specific `circular' case (Eq.~\ref{eq:circular_density}) and the highly generic `virial' case (Eq.~\ref{eq:virial_density}).
	
	\begin{figure}
		\includegraphics{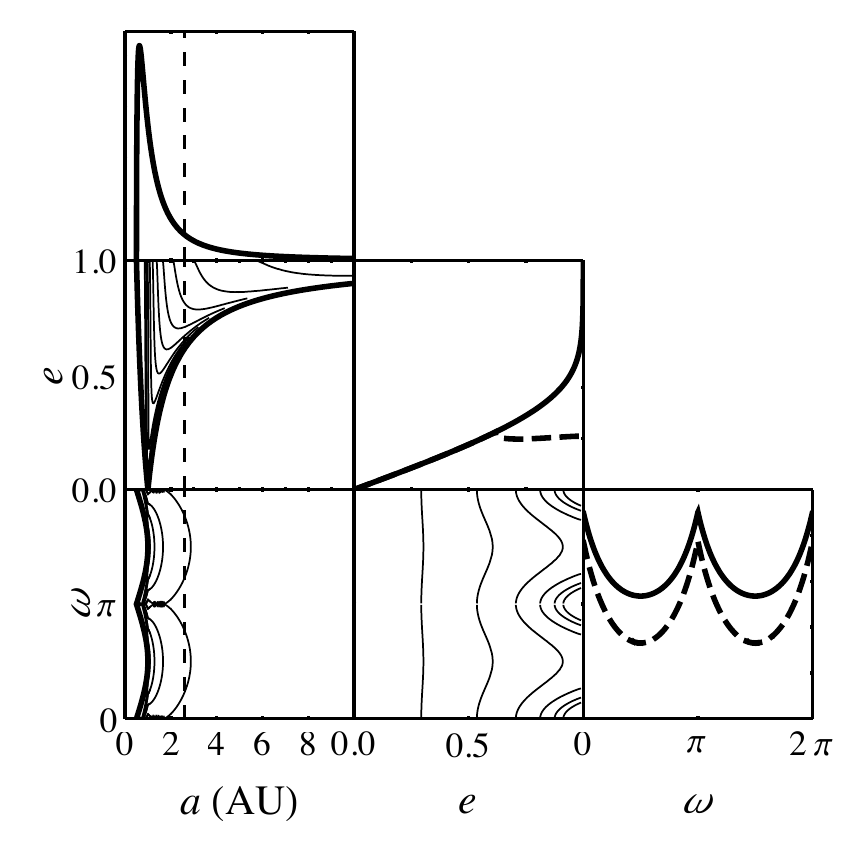}
		
		\caption{\label{fig:earthcrossingdist}Joint and marginal distributions of semimajor axis $a$, eccentricity $e$, and argument of perihelion $\omega$ for Earth-crossing orbits, assuming a primordial virialized distribution. The dashed line at $a=a_J/2\simeq 2.6\unit{AU}$ divides orbits that will never cross Jupiter's orbit from those that eventually evolve into Jupiter-crossing orbits. The marginal distributions of $e$ and $\omega$ include (solid) and exclude (dashed) orbits with $a>a_J/2$. The distribution of the inclination $i$ is uniform over $[0,\pi)$ and independent of $a$, $e$, and $\omega$ before interactions with Jupiter and Earth.}
	\end{figure}
	
	We will ultimately be interested in the orbits that intersect Earth. The (analytic) distribution of Earth-crossing DM orbits is shown in Fig.~\ref{fig:earthcrossingdist} in terms of three orbital elements: $a$, the semimajor axis; $e$, the eccentricity; and $\omega$, the argument of perihelion (the angle from the ascending node to perihelion). Two relevant features are the concentration of orbits with $a<1\unit{AU}$ and the broad distribution of $e$. (In contrast, the `circular' initial distribution would have all Earth-crossing orbits at $a=1\unit{AU}$, $e=0$, and $\omega$ undefined.) These distributions are taken after adiabatic contraction, but prior to any perturbations from the planets. We will see in Sec.~\ref{sec:jupiter} that all orbits with $a\gtrsim 2.6\unit{AU}$ are at immediate risk of ejection by Jupiter, and distributions excluding those orbits are plotted with dashed lines. However, further perturbations to be discussed in Sec.~\ref{sec:earthcrossing} will alter these distributions over the age of the Solar System.
	
	\section{Perturbations due to Jupiter\label{sec:jupiter}}
	
	Because we are considering a DM population that was established prior to the formation of the Solar System, we must determine how much---if any---will remain today, some 4.6\unit{Gyr} later. In particular, Jupiter has had a significant influence on the formation and evolution of the baryonic part of the Solar System, including shaping the structure of the asteroid belt via orbital resonances~\cite{kirkwood_gaps}. However, the mechanisms for clearing out some orbits crucially depends on non-gravitational interactions (namely, collisions) which DM mostly lacks. We will briefly argue that only one class of orbits---those with semimajor axis $a\gtrsim a_J/2 \simeq 2.6\unit{AU}$---are at risk of ejection from the Solar System by gravitational interaction with Jupiter. Others (e.g.~\cite{damour_lett,damour_prd}) have come to the same conclusion by various arguments; here we present a simple perturbative argument.
	\begin{itemize}
		\item A single close encounter with Jupiter can easily scatter a DM particle from bound to unbound, removing it from the population contributing to the density in the Solar System. Furthermore, in the range of semimajor axis $a\sim a_J$, a region of overlapping resonances results in chaotic evolution of orbits~\cite{chaos_solar_system}. So we must find the set of orbits which \emph{never} cross Jupiter's orbit, over the lifetime of the Solar System. It will suffice to consider the long-term behavior of the semimajor axis $a$.
		
		\item The evolution of the semimajor axis of the orbit is given by the osculating element differential equation, which has $\dot a\propto \pd{(\delta V)}{M_0}$ where $\delta V$ is the perturbation to the gravitational potential due to Jupiter and $M_0$ is the mean anomaly of the DM orbit at $t=0$~\cite{methods_celestial}. The first-order secular perturbation can be derived by averaging over the positions of the DM and Jupiter separately; $\expect{\delta V}$ can not depend on $M_0$, so $\expect{\dot a}=0$ to first order in $GM_J$. A more detailed calculation shows that $\dot a = 0$ holds to second order in secular perturbation theory~\cite{methods_celestial,tisserand}. Any third-order effect will not be highly significant over the lifetime of the Solar System.\footnote{A third-order suppression from the mass of Jupiter alone gives $\sim (M_\Sun/M_J)^3\unit{yr}\sim10^9\unit{yr}$ for a typical timescale; there will be additional powers of $a_J/a$ coming from the gravitational potential and its derivatives that further lengthen the timescale.}
		
		\item Secular perturbation theory can break down in two ways: resonance (violating the assumption of indepence of the position of DM and Jupiter) or close encounters (violating the smallness of the perturbation). Resonant orbits do typically have secular variation of $a$, but the semimajor axis and other elements typically \emph{librate} with a secularly long period and some finite amplitude. It is important to remember that orbital resonances are not like those of driven, undamped linear oscillators; since the resonant frequency is a function of the (perturbed) elements, the resonance is nonlinear in an important way, and its amplitude is self-limiting. (Ref.~\cite{henrard_resonance} has shown this analytically in a simplified model of low-inclination, nearly-circular orbits.)
		As long as the libration does not bring the DM into a close encounter with Jupiter, it will remain in its (librating) orbit over an arbitrarily long time. 
		
		\item The conclusion is that in the absence of close encounters, $a$ is effectively constant. Furthermore, a DM particle with a small enough $a$ will \emph{never} have a close encounter with Jupiter, as the aphelion $a(1+e)$ is bounded from above by $2a$. Although only certain orbits with a given $a$ (determined by the orbit's $e$ and $\omega$) will actually intersect Jupiter's orbit, $e$ and $\omega$ vary secularly at leading order, and a typical orbit with $a\gtrsim a_J/2$ will `sweep through' Jupiter's orbit many times over the lifetime of the Solar System.
	\end{itemize}
	
	While all orbits with $a\lesssim a_J/2$ are safe from expulsion by Jupiter over the lifetime of the Solar System, we must consider all orbits with $a\gtrsim a_J/2$ as potentially lost. In the conservative `virial' distribution considered in Sec.~\ref{sec:distribution}, 22.4\% of the total Earth-crossing density is at risk of crossing Jupiter's orbit, before considering perturbations from Earth and other planets.
	
	\subsection{Secular variation in other elements\label{sec:lidovkozai}}
	
	Although we have argued that DM orbits with $a<a_J/2$ are long-lived with secularly constant $a$, it is worth discussing the nontrivial secular changes in the other orbital elements. The secular variation of $\omega$ is (part of) the well-known precession of perihelion, which is the most important effect for nearly-circular ecliptic orbits. The secular variation of eccentric and/or inclined orbits is much more dramatic: $e$ and $i$ undergo Lidov--Kozai oscillations~\cite{lidov,kozai} induced by Jupiter. Lidov--Kozai oscillations exchange inclination for eccentricity while keeping $\sqrt{1-e^2}\cos i$ constant. As the inclinations of primordial DM orbits are expected to be uniformly distributed, and only a modest inclination is required to achieve significant eccentricity, oscillations will eventually drive much of the phase space to be highly eccentric as well. For this reason, assuming a distribution of highly circular orbits (such as the sharply-peaked Eq.~\ref{eq:circular_density}) is unrealistic.
	
	Another practical upshot of the Lidov--Kozai oscillations of DM orbits is that individual Earth-crossing orbits do not remain Earth-crossing for very many orbits per oscillation; this will be important in Sec.~\ref{sec:earthcrossing} for preventing Earth from clearing its entire orbit of DM.
	
	\section{Earth-crossing Orbits\label{sec:earthcrossing}}
	
	\begin{figure*}
		\includegraphics{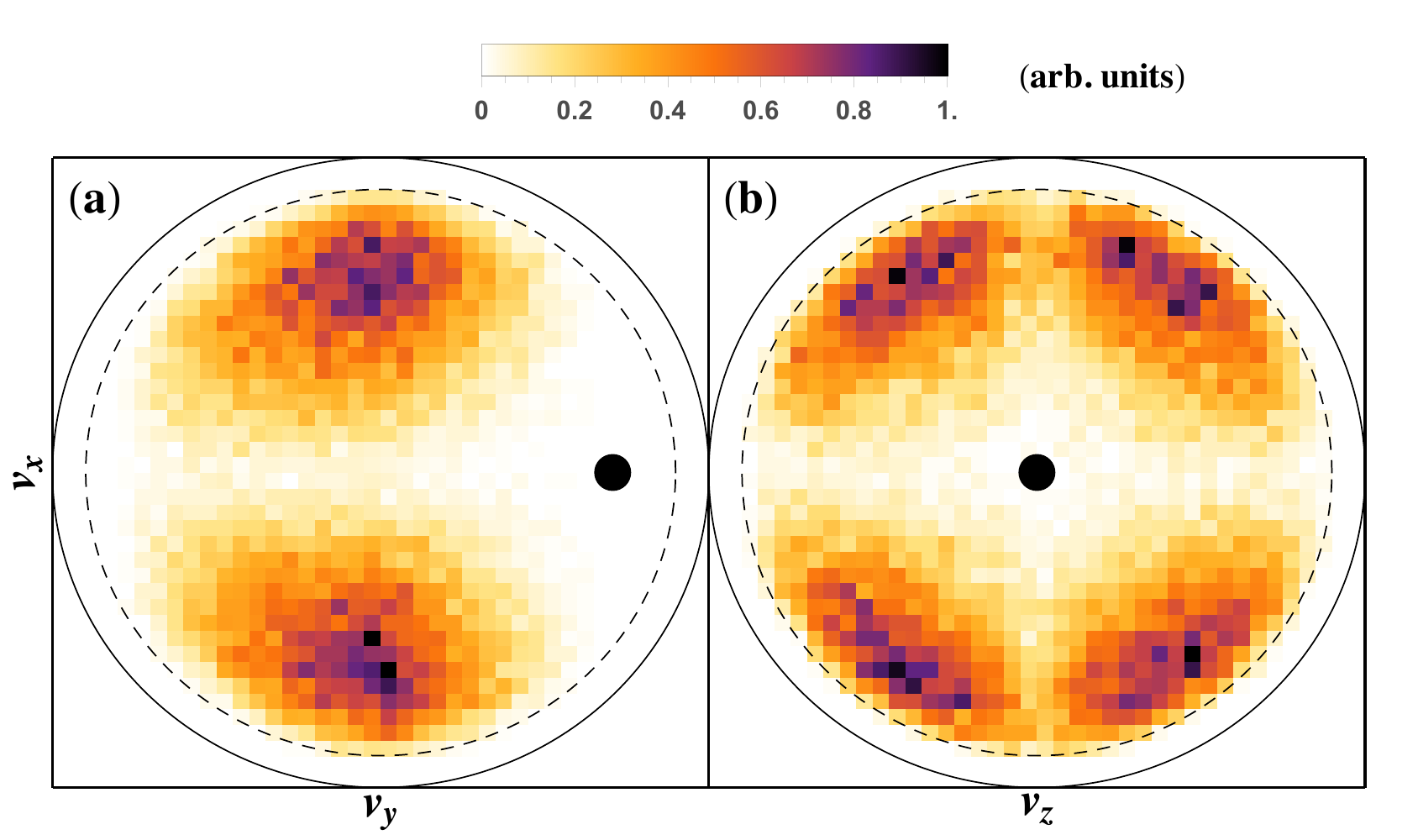}
		 
		\caption{\label{fig:diffusion}Velocity-space distribution of Earth-crossing orbits after 4.6\unit{Gyr} of simulated gravitational diffusion. Darker (more purple) squares have greater projected phase space density. The two panels are projections (not slices) of the 3D velocity space in two different planes. The axes are oriented such that $x$ is radial (up the page is toward the Sun), $y$ is tangential (right is along Earth's orbit), and $z$ is perpendicular to the ecliptic (right is towards the north ecliptic pole). The black dot is Earth's velocity, which here is purely in the $y$ direction. Orbits outside the dashed circle have $a>a_J/2$ and are eventually ejected by Jupiter, while orbits outside the solid circle are above Solar escape velocity. Most of the orbits that remain are moderately eccentric and moderately inclined (i.e.~have nonzero $v_x$ and $v_z$), and there is a slight bias toward retrograde orbits.}
	\end{figure*}
	
	Because we are primarily interested in directly detecting DM particles in Earth-based experiments, it is necessary to consider the DM orbits that make (maximally) close encounters with Earth. In contrast with Jupiter, single close encounters with Earth deflect particles by only a small angle ($\lesssim 0.1\unit{rad}$ for typical incoming speeds) and are ineffective at transitioning most bound orbits to unbound orbits. Naively, a particle in an Earth-crossing orbit will continue crossing Earth's orbit once per period, providing repeated attempts at scattering, and random-walking around the accessible phase space. This naive diffusion process overestimates the scattering of the orbit due to the secular perturbations from Jupiter: Lidov--Kozai oscillations (or, indeed, normal perihelion precession) will modify the orbit over time, such that it will `sweep through' Earth's orbit twice per secular oscillation period. As long as the secular oscillation period is $\ll \unit*{Gyr}$, the naive diffusion rate will be reduced by the fraction of time the orbit spends crossing Earth's orbit. The accumulated effect of many close encounters with planets has been termed ``gravitational diffusion'' and studied in detail by Gould~\cite{gould_diffusion}. Although Gould does not account for the full Lidov--Kozai behavior of inclined orbits, he does account for perihelion precession~\cite{gould_paper2}, which is at most $O(1)$ different from the full treatment.
	
	Figure~3 of Ref.~\cite{gould_diffusion} depicts the regions of the Earth-crossing phase space which are effectively equilibrated with the unbound phase space (the Galactic DM halo). As previous work has shown~\cite{peter_dmss_iii,edsjo_comments}, the resulting particle density in bound orbits is very low due to the (relatively) small volume in velocity space. The regions of phase space which are not equilibrated are labeled `unfilled,' but if they were filled initially (e.g.~by adiabatic contraction, or by particle production in the Sun~\cite{stellar_basins}) they may remain filled today, potentially at a detectably large particle density.
	
	However, the fact that these orbits remain unfilled in Ref.~\cite{gould_diffusion} does \emph{not quite} imply that they would all remain filled if populated initially. In fact, some of the unfilled phase space has a relatively short diffusion time; it remains unfilled because it is separated from the unbound phase space by a region of long diffusion time. Furthermore, the `long' diffusion timescale is actually not much longer than the age of the Solar System, so some evaporation will occur as Earth `upscatters' DM into Jupiter-crossing orbits.
	
	To quantify the diffusive loss of DM and complete our argument that (at least some) Solar halo DM is long-lived, we simulate the random walk of gravitational diffusion. A sample of $10^5$ orbits, chosen from the virial initial distribution (Sec.~\ref{sec:distribution}), was evolved over 4.6\unit{Gyr} of close encounters with Earth using the diffusion rate calculated in Ref.~\cite{gould_paper2}. Each orbit's evolution is divided into small enough timesteps so that the maximum scattering angle is $0.1\unit{rad}$, roughly the maximum that a single encounter with Earth can achieve. Orbits that reach $a\ge a_J/2$ at any timestep are removed. Certain orbits which are very close to Earth in velocity space are removed manually, which accounts for the small effect of Venus diffusing those orbits into ones that Earth can scatter more effectively~\cite{gould_diffusion}.
	
	The resulting velocity space distribution is summarized graphically in Fig.~\ref{fig:diffusion}. The fraction of all initially Earth-crossing orbits that remain after 4.6\unit{Gyr} is 24.4\%. While there is significant `evaporation' under the combined action of Earth, Jupiter, and Venus, a substantial portion of the phase space is long-lived. The most notable features of the final distribution are the clearing of the area near Earth's velocity (consistent with Fig.~3 of Ref.~\cite{gould_diffusion}) and the concentration of orbits that are moderately inclined, moderately eccentric, and retrograde. Zero-inclination orbits intersect Earth's orbit over a larger area ($\sim 1/\sin i$), and zero-eccentricity orbits remain Earth-crossing even while undergoing perihelion precession, so these types of orbits are cleared out effectively. Prograde orbits have lower speeds relative to Earth, thus larger scattering angles, and are also closer to Solar escape velocity, thus easier to eject.
	
	Combining the results of our simulation (a 24.4\% diffusion survival rate) with Eq.~\ref{eq:virial_density}, we obtain the final density of Solar halo DM at Earth, today:
	\begin{align}
		\rho_\Earth &\simeq 2\times 10^6 \rho_i \left(\frac{R_i}{\unit*{ly}}\right)^{3/2} \label{eq:final_from_density} \\
		&\simeq 0.06\unit{\frac{GeV}{cm^3}}\left(\frac{M_{\rm DM}}{10^{-10}M_\Sun}\right) \left(\frac{R_i}{\unit*{ly}}\right)^{-3/2} \label{eq:final_from_mass}
	\end{align}
	where $M_{\rm DM}$ is the total mass of DM bound to the gas eventually becoming the Sun. (In Sec.~\ref{sec:introduction}, we argued that one part in $10^{10}$ of DM would result in a density competitive with the Galactic halo; Equation~\ref{eq:final_from_mass} gives the more accurate answer accounting for gravitational diffusion over the age of the Solar System.)
	
	\section{Observable Signatures\label{sec:detection}}
	
	
	\begin{figure}
		\includegraphics{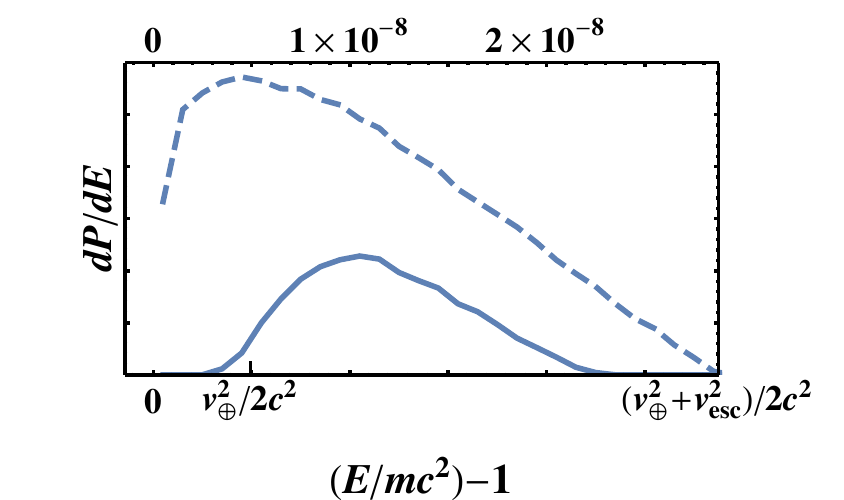}
		
		\caption{\label{fig:spectrum}Distribution of specific kinetic energy of Earth-crossing Solar halo DM in Earth's reference frame; equivalently, the power spectral density of a signal from ultralight DM that would appear in (e.g.) an axion or dark photon detector. The left endpoint is at the rest energy (or frequency) of the DM particle, while the right endpoint is the energy of a DM particle with Solar escape velocity moving opposite to the Earth. The solid (dashed) curve is after (before) 4.6\unit{Gyr} of gravitational diffusion, and the graphs are both normalized to the same \emph{initial} density, so that the area under the solid curve is $24.4\%$ of the dashed curve. Gravitational diffusion has depleted the extreme ends of the spectrum: the left end is strongly diffused by Earth, and the right end is strongly diffused by Jupiter.
		The corresponding spectrum of an equal density of Galactic halo DM is roughly 100 times wider, but 100 times lower, than that of the Figure.}
	\end{figure}
	
	While we make no attempt to predict the actual density of Solar halo DM at Earth, we consider it a promising experimental target for ultralight DM experiments. Importantly, a candidate signal would have several distinctive features which could rule out Galactic halo explanations. These features are similar to (but distinguishable from) the signatures of nonvirialized cold DM `streams'~\cite{caustic_rings} which are targets of, for example, ADMX's high resolution (HR) channel~\cite{admx_hr_new,Hoskins:2011iv}. The existence of such nonvirialized streams in the Galaxy depends on the detailed self-interactions of the DM particle~\cite{axion_bec}; our proposal provides a new motivation for these searches that does not depend on those self-interactions. For a review of several current proposals to detect axions and axion-like particles, see Ref.~\cite{graham_review}.
	
	Since ultralight DM experiments are typically coherent detectors, the narrow velocity dispersion of DM bound to the Sun will result in a $\sim 100\times$ sharper signal in frequency space at a given density.  Resonant cavities with high enough $Q$ and long enough integration times can build up a larger signal power at a given DM density.  However, even with cavity $Q$ and integration time targeted for Galactic halo searches, Solar halo DM can have a better signal-to-noise ratio at a given DM density if the frequency resolution of the data collection pipeline is sufficiently good. For example, ADMX performs their HR channel search in parallel with the standard axion DM search: The same resonant cavity power is analyzed with a higher frequency resolution to look for narrow spectral features, resulting in better limits on density for low-velocity-dispersion DM~\cite{admx_hr_new,Hoskins:2011iv}.
	
	More advanced searches (presumably follow-up experiments) could verify detailed properties of the frequency spectrum. The lineshape of the signal, determined by the energy spectrum in the Earth frame, is shown in Fig.~\ref{fig:spectrum}. The fractional width of the signal is $\Delta f/f \sim v_\Earth^2/c^2 \sim 10^{-8}$. Unlike a Galactic DM signal, there is not a significant annual modulation in the Earth-frame DM energy, since the DM is co-orbiting with the Earth. There are two effects due to Earth's eccentricity $e\sim 0.017$: an $O(e)$ variation in the already-small width of the line (due to the changing relative speeds of Earth and the DM as they speed up and slow down in orbit), and an $O(e)$ variation in the amplitude, due to moving through the radial DM density profile. For a density profile $\rho(r)\propto r^{-n}$, the fractional change in DM density between aphelion and perihelion is
	\begin{equation} \frac{\Delta \rho}{\rho} \sim 2en. \label{eq:densityvariation} \end{equation}
	The virialized initial distribution ($n=3/2$) predicts a $5\%$ variation in the DM density over half a year, a small but potentially important confirmatory signal.
	The phases of the annual modulations are fixed; the largest width and amplitude of the signal must occur near perihelion, around January 4. Compared to annual modulation searches for ultralight Galactic halo DM, the roles of frequency and amplitude are essentially swapped, and the phase of the signal is shifted from a maximum around June 1, when Earth is comoving with the Sun around the galaxy~\cite{annual_mod_review}.
	
	The interesting phase space structure of gravitationally diffused DM (Fig.~\ref{fig:diffusion}) could be explored with direction-sensitive DM detectors; for example, the CASPEr-Wind experiment relies on a velocity-sensitive coupling of axions to nucleons~\cite{casper_proposal}. However, the low velocity of Solar halo axions in the Earth frame makes such an experiment highly challenging.
	
	Finally, we mention potential applications of our work to non-ultralight DM. WIMPs have been the traditional focus of work on DM in the Solar System~\cite{peter_dmss_i,peter_dmss_iii,edsjo_diffusion,edsjo_comments}, but their nuclear scattering cross sections typically lead to depletion through Solar or terrestrial capture. As WIMP searches exclude cross sections down to lower and lower values, it may be that even WIMPs survive for long periods of time in the Solar System. Unfortunately, WIMPs with Solar orbit velocities have significantly lower kinetic energy to excite detectors via scattering, and most events will fall below threshold. Unless the Solar halo density is several orders of magnitude larger than the Galactic density (which is not impossible, see Fig.~\ref{fig:densitylimits}), WIMP detectors are unlikely to see any signal of a Solar halo.
	
	\section{Discussion\label{sec:discussion}}
	
	We have shown that, if some DM is initially bound to the gas cloud that eventually became the Sun, a significant fraction would remain bound in Earth-crossing orbits today, after accounting for 4.6\unit{Gyr} of gravitational perturbations from Jupiter, Earth, and Venus. Previous work that found negligible amounts assumed either zero initial abundance (a possible, if highly special, initial condition) or depletion of WIMPs in the Sun and Earth via nuclear scattering. Ultralight DM candidates in Solar orbit are simultaneously less likely to suffer similar depletion (though this statement is model-dependent) while also more promising to detect. A Solar halo population of ultralight DM would prove an especially interesting target for coherent direct detection experiments due to the low velocity dispersion and distinctive annual modulation.
	
	Although we do not predict a specific density of Solar halo DM at Earth, very small initial abundances (a part in $10^{10}$) give rise to local densities comparable to that of the Galactic halo; furthermore, planetary motion limits on the local population of DM allow densities several orders of magnitude higher. Therefore, we find that a detectable abundance is plausible, though we do not set any limits from non-detection.
	
	While not the focus of this paper, our results on longevity may have important consequences on non-DM particles in the Solar System; see, for example, Ref.~\cite{stellar_basins} in which Solar axions are produced into bound orbits and then diffused into a Solar halo. Follow-up simulations are planned, accounting for the different initial conditions of this scenario.
	
	The biggest question we leave for future theoretical work is the question of the initial abundance. It may be necessary to follow bound DM through multiple generations of star formation in order to properly predict this value, thereby allowing exclusion of (Galactic) DM models by non-detection of a Solar population. In the interim we hope for DM direct detection experiments to begin (or continue) searching for low-dispersion signals. If a detection is made, we may learn as much about the history of our Solar System as we do about the nature of dark matter.
	
	\begin{acknowledgments}
		We are thankful to Joakim Edsj\"o for productive discussions during the development of this work. We especially appreciate Ken Van Tilburg for discussions and comments on a draft of this manuscript.
		
		TDW thanks Marc Robbins for collaboration on an early version of this idea, Stanford Institute for Theoretical Physics for hospitality during that time, as well as Peter W.~Graham for helpful early discussions.
		
		NBA and TDW acknowledge the support of the Ronald E.~McNair Postbaccalaurate Program grant. AP and TDW acknowledge the support of the TruScholars summer research scholarship.
		
		Some figures for this paper have been created using the SciDraw scientific figure preparation system~\cite{scidraw}.
		
		The contents of this article were developed under a grant from the U.S.~Department of Education. However, those contents do not necessarily represent the policy of the Department of Education, and you should not assume endorsement by the U.S.~Federal Government.
		
	\end{acknowledgments}
	
	\appendix
	
	\section{Orbital Elements\label{app:elements}}
	\begin{table}[ht]
		\caption{Symbols for Keplerian elements used in this paper.\label{tab:elements}}
		\begin{ruledtabular}
		\begin{tabular*}{\linewidth}{cl}
			Symbol & Element \\ \hline
			$a$ & semi-major axis \\
			$e$ & eccentricity \\
			$i$ & inclination \\
			$\omega$ & argument of perihelion \\
			$\Omega$ & longitude of ascending node \\
			$M_0$ & mean anomaly at epoch
		\end{tabular*}
		\end{ruledtabular}
	\end{table}
	
	Table~\ref{tab:elements} lists the symbols and definitions of the Keplerian orbital elements used throughout this paper. All six elements are constant for unperturbed orbits. When orbits are perturbed, these symbols refer to the osculating elements: those computed from a tangent Keplerian orbit. The osculating elements are no longer constant.
	
	A Solar System orbit can be constructed geometrically from the elements by taking an ellipse (semimajor axis $a$, eccentricity $e$) in the $xy$ plane with its focus (the Sun) at the origin and the perihelion along the $+x$ axis. Then, rotate by $\omega$ about the $z$ axis, $i$ about the $x$ axis, and finally by $\Omega$ about the $z$ axis. The time-dependent position along an orbit is not essential for this work.
	
	Of special interest is whether or not an orbit intersects that of Earth (or another planet). We will assume a circular, ecliptic orbit for Earth with $a_\Earth=1\unit{AU}$. The $i=0$ case (coplanar orbit) is special: as long as $a(1-e)<a_\Earth<a(1+e)$, the orbits will cross. If $i\neq 0$ then the orbits will intersect if and only if either the ascending or descending node intersects Earth's orbit, i.e.~has radius $a_\Earth$. The radii of the ascending and descending nodes are given by
	\begin{align}
		r_{AN,DN} &= \frac{a(1-e^2)}{1\pm e\cos\omega}.
	\end{align}
	However, for most orbits, $\omega$ precesses significantly due to perturbations. As a result, $r_{AN,DN}$ sweep through all values $a(1-e)<r_{AN,DN}<a(1+e)$, reducing to the same general result as the $i=0$ case, but with a suppressed encounter frequency.
	
	\section{Adiabatic Invariants\label{app:invariants}}
	
	As discussed in Sec.~\ref{sec:distribution}, we compute the phase space density of Solar halo DM after adiabatic contraction using the action variables in spherical coordinates. Both the initial and final Hamiltonians take the form 
	\begin{equation}
		H=\frac{p_r^2}{2}+\frac{\ell_\theta^2}{2r^2}+\frac{\ell_\phi^2}{2r^2\sin^2\theta}+V(r)
	\end{equation}
	where $(r,\theta,\phi)$ are spherical coordinates and $(p_r,\ell_\theta,\ell_\phi)$ are their conjugate momenta. $\ell_\phi$, $L\equiv\sqrt{l_\theta^2+\ell_\phi^2/\sin^2\theta}$, and $E\equiv H$ are conserved quantities. The action variables are then defined by
	\begin{equation}
		J_i = \oint p_i\,dq_i\qquad\textrm{(no sum)}
	\end{equation}
	where $i\in \{r,\theta,\phi\}$ and the integral is taken over a full period of the particle's trajectory. Without specifying $V(r)$ we find
	\begin{align}
		J_\phi &= 2\pi\ell_\phi \label{eq:jphi}\\
		J_\theta &= 2\pi(L-\ell_\phi) \label{eq:jtheta}
	\end{align}
	in terms of conserved quantities. The form of $J_r$ depends on the potential. For simple harmonic potential $V_{\rm SHO}(r)=kr^2/2$,
	\begin{equation}
		J_r = -\pi L + \frac{\pi}{\sqrt{k}}E, \label{eq:jrsho}
	\end{equation}
	while for the Kepler potential $V_{\rm K}(r)=-\mu/r$, we obtain
	\begin{equation}
		J_r = -2\pi L + \frac{2\pi\mu}{\sqrt{-2E}}. \label{eq:jrk}
	\end{equation}

	When either potential is chosen and remains time-independent, the action variables are also constant in time. (This is a general result of Hamiltonian mechanics in action--angle variables, but it can also be seen explicitly from their formulas.) Furthermore, when the potential slowly changes from $V_{\rm SHO}$ to $V_{\rm K}$, the originally conserved quantities are not necessarily conserved, but the action variables remain constant in the adiabatic limit; in particular, we can use the adiabatic invariance of the $J_i$ to determine the final values of $\ell_\phi$, $L$, and $E$ and therefore the final orbits.
	
	Equations~\ref{eq:jphi} and~\ref{eq:jtheta} imply that $\ell_\phi$ and $L$ are both adiabatically constant even as the potential varies. However, $E$ is not adiabatically constant; it can be solved for explicitly by equating Eqs.~\ref{eq:jrsho} and~\ref{eq:jrk}.

	\bibliographystyle{apsrev4-2}
	\nocite{APSREV42CONTROL}
	\bibliography{dmss}
	
\end{document}